\documentclass[10pt,conference]{IEEEtran}
\IEEEoverridecommandlockouts
\usepackage{cite}
\usepackage{amsmath,amssymb,amsfonts}
\usepackage{algorithmic}
\usepackage{graphicx}
\usepackage{textcomp}
\usepackage{xcolor}
\usepackage{soul}
\usepackage{array}
\usepackage{float}
\def\BibTeX{{\rm B\kern-.05em{\sc i\kern-.025em b}\kern-.08em
    T\kern-.1667em\lower.7ex\hbox{E}\kern-.125emX}}
\begin{document}

\title{Addressing Age-Related Accessibility Needs of Senior Users Through Model-Driven Engineering \\
}

\author{\IEEEauthorblockN{Shavindra Wickramathilaka}
\IEEEauthorblockA{\textit{Software Systems and Cybersecurity} \\
\textit{Monash University}\\
Melbourne, Australia \\
Shavindra.Wickramathilaka@monash.edu}
\and
\IEEEauthorblockN{Ingo Mueller}
\IEEEauthorblockA{\textit{Software Systems and Cybersecurity} \\
\textit{Monash University}\\
Melbourne, Australia \\
Ingo.Mueller@monash.edu}
}

\maketitle

\begin{abstract}
One of the main reasons that cause seniors to face accessibility barriers when trying to use software applications is that the age-related user interface (UI) needs of seniors (e.g., physical and cognitive limitations) are not properly addressed in software user interfaces. The existing literature proposes model-driven engineering based UI adaptations as a prominent solution for this phenomenon. But in our exploration into the domain, we identified that the existing work lacks comprehensiveness when it comes to integrating accessibility into software modelling tools and methods when compared to a well-recognised accessibility standard such as the Web Content Accessibility Guidelines (WCAG). Thus in this paper, we outline a research roadmap that aims to use WCAG as a reference framework to design domain-specific languages that model the diverse accessibility scenarios of senior users via user context information and UI adaptation rules modelling so that they meet the accessibility standards specified in WCAG. 
  
\end{abstract}


\section{Introduction} \label{Intro}

According to the World Health Organisation (WHO), one in six people will be sixty years of age or older by 2030 \cite{b1}. Despite this trend, developers neglect to cover the software accessibility needs of the senior target audiences resulting in digital exclusionary practices \cite{b22}. Studies such as \cite{b19, b24, b22} have shown that several of the key challenges that cause a digital divide for seniors can be attributed to user interfaces (UI). These all too often do not reflect the diversity of age-related physical and cognitive challenges experienced by seniors such as vision and hearing impairments, psycho-motor disabilities, and memory limitations. As a consequence, the interactions of seniors with their software applications are hampered by accessibility barriers \cite{b7}.

Therefore it is vital that software developers design specifically for seniors and address their accessibility needs. However, doing this is not a trivial matter, due to the inherent diversity and variability of age-related needs of each individual senior user. In other words, it is not always possible to have complete knowledge of all accessibility requirements at design-time. Furthermore, manually building accessibility into a UI right from the start can be a complex problem and cost ineffective according to \cite{b21}.

An important concept that may aid in addressing this is Model-Driven Engineering (MDE) \cite{b4}. MDE considers software models as the primary artefacts for the generation of application code with minimum to no handwritten coding involved \cite{b4}. As described in multiple studies \cite{b7, b12, b13, b16} that we present in Section \ref{RW}, Domain-Specific Language (DSL) software models enable UI designers and even domain experts to comprehensively, and formally express age-related UI needs. These DSL models can be tailored to the needs of our end users at design-time and input into an MDE code generation process to produce adaptable and accessible UIs.

Aside from MDE’s automated code generation capabilities, where it really helps us to address this issue is when we model UI adaptation capabilities as part of an MDE process. As age-related UI needs are variable and diverse, we can at best capture a snapshot of a senior’s needs, e.g. in the form of personal, device and environment related context information, through models at design-time. However, context dimensions can change at any time and if so, we are looking at a context scenario outside of what has been modelled. One way to respond to such context changes is the integration of UI adaptation techniques into the MDE process at both run and design-time to adapt our models accordingly and to generate an adapted UI instance. This is, according to a 2014 literature survey \cite{b21}, a prominent approach for addressing context variability of UIs.

In this paper, we explore the state-of-the-art of addressing the accessibility needs of seniors through adaptable MDE-driven methods. We firstly report on a literature survey in Section \ref{RW} that demonstrates that the topic has attracted attention in recent years. However, most approaches are incomplete when compared against generally accepted accessibility development standards such as the World Wide Web Consortium’s (W3C) WCAG \cite{b2}. While some approaches incorporate WCAG and similarly relevant standards into their MDE tools, our in-depth examination of selected papers suggests that multiple improvements can be made to the capabilities of existing modelling approaches and adaptation techniques (in Section \ref{RG}). Secondly, we present our research roadmap in section \ref{RR} for the creation and evaluation of a Model-driven accessible-adaptive UI modelling and code generation tool that will close the gap for selected accessibility needs.

\section{Background}\label{RC}

\subsection{Accessibility}

In the context of information technology, the term 'accessibility' refers to the \textit{"operational suitability of both hardware and software for \textbf{any} potential user. "} \cite{b28}. 

\begin{itemize}
    \item \textit{Web Content Accessibility Guideline (WCAG)}:

    WCAG provides a well-known set of recommendations to make web content more accessible to people with disabilities such as hearing impaired, vision impaired, and cognitively impaired. The standard is structured into a hierarchy of principles, guidelines, success criteria, and advisory techniques \cite{b2}.
\end{itemize}

\subsection{User interface adaptation}
UI adaptation can be defined as the ability of a system or a user to adapt the UI \cite{b7}. Example UI adaptations are layout, navigation, styling, and I/O modality changes. 


\begin{itemize}
    \item \textit{Design-time UI adaptation}: 
    refers to the UI design process that takes place during UI development. Developers design and tailor UI models according to known and static CoU information and the resulting models can then be used to generate the Final UI.
        
    \item \textit{Run-time UI adaptation}:
    takes place while the application's final UI is running. These adaptations can happen due to dynamically changing CoU scenarios or user-adaptable configurations.

    \item \textit{Context-of-Use (CoU)}:     
    Calvary et al., \cite{b3} define the CoU as an n-tuple that comprises the following entities: 1) \textbf{user} of the system and their attributes (e.g. impairments and language preferences), 2) \textbf{platform}: hardware and software components that the users use to interact with the system (e.g. device type and screen dimensions), and 3) \textbf{environment} in which interactions between the user and the system occur (e.g. ambient noise and weather). When adapting the UI, the CoU can be used to model user-centric accessibility scenarios.
\end{itemize}

\subsection{Model-driven Engineering (MDE)}
MDE is a software development methodology where software models are considered to be the primary artefact of the software development process. When the models are passed into an MDE process, they are transformed into code artefacts automatically. \cite{b4}

\begin{itemize}
    \item \textit{Domain-specific language (DSL)}:

    A DSL is a software design and/or development language that is capable of expressing application domain-specific concepts. The expressive power and the notations of a DSL enable analysts and domain experts to communicate the domain requirements to the developers. \cite{b4} 

    \item \textit{Abstract UI model (AUI)}:
    A model that renders UI domain concepts explained in a DSL, independent from the target platform or modes of interaction. A similar term for AUI model is 'platform independent UI model'. \cite{b3}

    \item \textit{Concrete UI model (CUI)}:
    CUI renders UI domain concepts in a platform and interaction mode-dependent expression. In MDE, the CUI model is the result of transforming AUI into a platform-specific model that explicitly defines the look and feel of the final UI. \cite{b3}

    \item \textit{Final UI (FUI)}:
    In MDE, FUI is the source code that is generated from transforming the CUI model. It can then be interpreted/compiled into an executable UI. (Some approaches require handwritten additions from developers to make the generated code executable.) \cite{b3}
\end{itemize}

\section{Related Work}\label{RW}

In this section, we briefly review key literature on adaptive UIs and MDE-based adaptive UIs. The literature review selection criteria were: 1) approaches using MDE-based methodologies, 2) approaches addressing UI adaptations, and 3) recent works being published after 2010.



        

    \subsection{Design-time UI adaptive only MDE approaches}\label{DT}

    In a recent (2022) study, Braham et al., \cite{b8} present an MDE approach for generating adaptive UIs at design-time to cope with heterogeneous CoU scenarios in pervasive computing applications. To achieve this, they use different ontologies to model UI, CoU scenarios, and adaptation rules. The latter two models adapt the UI model depending on the accessibility scenario. The paper presents several limitations: 1) The Context Ontology metamodel does not cover device-related context information such as platform, device, network, and screen dimensions. Not enough information is available to determine the user and environmental context modelling capabilities of the Ontology model. 2) The given adaptation rules metamodel is not sufficient to model conditionally compound adaptation rules and modality changes (e.g., screen readers, text-to-speech). Furthermore, UI layout adaptations can be further expanded to include styling changes such as border, margin, padding, and font.    

    ‌Bendaly Hlaoui et al., \cite{b7} present an MDE approach that is only capable of design-time UI adaptations. However, their CoU and UI adaptation modelling approaches are comparatively more mature than the previous study, and their work is also motivated by the UI adaptation needs of disabled users. The CoU scenarios are modelled based on a novel ontology that provides disability, user, device, and environmental context information. The elements from the ontology instance model are then annotated into the non-adapted abstract UI model (AUI) of the software to generate an adapted AUI model. From a user study evaluation with blind, low-vision, and deaf participants, the authors discovered that their adapted UIs were found to be less satisfactory to participants with less computing platform usage experience. While this finding has not been discussed further in detail, the authors do suggest user configuration options based on user computer expertise. Therefore a key opportunity to extend the approach is by providing run-time UI configuration options to the end users. The level of configuration options provided to users would depend on their computer experience level.

    In another study, Bacha et al. \cite{b5} propose an MDE approach to personalise UIs by adapting them at design-time. This is achieved via three modelling artefacts: 1) A Context model to capture Context-of-Use scenarios; 2) A Domain Ontology model to capture and model the intended application’s domain information; 3) A Business Process model to express tasks and information flow between them. The mappings between these three models allow the designers to generate an Abstract UI (AUI) model that is subsequently transformed into the Final UI in an MDE process. A limitation of their study is that while the authors emphasise UI personalisation, compliance with UI accessibility standards (e.g., WCAG) is not considered.

    Distinct from the previous  works, Park and Lee \cite{b6} propose a Requirements Engineering (RE) method that introduces the three abstraction levels of MDE when documenting adaptive UI requirements. The method contains a 3-layered RE notation: 1) Abstract UI requirements: to express a high-level domain/platform independent adaptive UI goal/requirement. 2) Concrete UI Requirements: to further elaborate the AUI requirement to include domain/platform-specific functional and qualitative UI requirements. 3) FUI requirements: detail the requirement further to include implementation details such as trigger rules, sensor type, and usability metrics. The lack of a user study with RE practitioners limits the reader’s ability to comprehend the effectiveness of the proposed conceptual method. A further limitation of the method is the lack of discussion beyond the requirement engineering phase. The proposed adaptive UI requirements capturing notation has to be integrated with a model-driven process to transform the defined requirements into an executable UI.

    \subsection{Run-time UI adaptation capable MDE approaches}\label{RT}

    Yigitbas et al.’s \cite{b16} 2020 work proposes two novel DSLs (ContextML and AdaptML) that are integrated with the Interactive Flow Modelling Language (IFML). IFML is an OMG standard that supports the platform-independent definition of UIs \cite{b17}, and this work enables the generation of software that can automatically adapt the user interfaces at run-time. ContextML models CoU scenarios while AdaptML models adaptation rules that reference both CoU elements in the context model and an abstract UI model specified in IFML. ContextML is limited in modelling user context elements due to its metamodel. For example, ContextML is not able to define a CoU scenario based on the user’s proficiency or familiarity with a language or the accents/dialects used in its voice output modality. Similarly, AdaptML is also limited in its adaptation capabilities. For example, several WCAG guidelines such as text alternatives, time-based media, and timed activities (i.e., timeouts, interruptions, re-authentications) are not covered in the AdaptML metamodel.

    Ghaibi et al. \cite{b12} propose an MDE approach for generating CoU-sensitive and adaptive UIs. At design-time, developers are provided with a graphical MDE toolkit to manage CoU parameters, manage adaptation rules, and generate tailored UIs. At run-time, users also have the freedom to create their own UI adaptation rules for each CoU scenario, manage them, and generate adapted UI instances. The paper presents several limitations: 1) CoU metamodel is shallow and does not allow in-depth and comprehensive CoU scenario modelling. For example, the prototype example indicates a limited set of context parameters being modelled but parameters such as language, age, exact screen dimensions, ambient light/noise, and speech-to-text modality are not considered. 2) Metamodel does not support UI adaptation logic such as compound adaptation rules and layout-related UI adaptations (i.e., font type, margin, and border). This limits the opportunities to comply with UI accessibility guidelines presented in WCAG.

    Akiki et al. \cite{b13} present a Role-Based UI adaptation mechanism for addressing context-of-use (CoU) scenarios. The MDE metamodel of the approach only supports graphical/layout UI adaptations whereas multi-modality (auditory and tactile) features such as text-to-speech, speech recognition, and keyboard-only navigation are not covered. UI adaptations themselves are achieved via adopting elements of the Role Based Access Controls (RBAC) \cite{b14}. Here, UI elements and UI adaptation rules are considered accessibility resources and allocated to user accounts as ‘roles’. At design-time, a user’s account is assigned roles that are associated with CoU information such as disabilities and culture. But since some context information (e.g., ambient brightness and noise level) are only known at run-time, the roles associated with the latter group of context information are dynamically allocated to a user account once a UI adaptation workflow is triggered. The approach can be extended by the inclusion of multi-modality features to improve the accessibility of the adapted UIs to user groups such as the Vision-impaired and seniors. Another opportunity is the addition of navigational adaptation support for the MDE UI metamodel ensuring further WCAG compliance.

    A further extension of this work by Akiki \cite{b15}, who proposes Contextual Help for Adaptive INterfaces (CHAIN). This provides a complementary approach to model-driven adaptive UI systems for developing context-aware UI help that covers different UI adaptation scenarios. UI help models are transformed into CHAINXML documents and are kept separately from the MDE UI models. The integration between help UI elements and running UI elements is based on mappings between them. This allows the running UI to maintain its help features even when navigational, layout, presentation, and widget-based changes (resizing, relocating, and changing UI widgets) occur. A limitation of the study is that the contextual help UI overlays are not fully aware of how the software UI adapts itself. An example situation that can occur because of this limitation is, when the UI adapts itself to a vocal-only I/O modality on run-time, the CHAIN help UI cannot switch to vocal-only help.
    
    Bongartz et al. \cite{b9} propose an approach to supporting model-driven adaptive UIs in work environments, with an emphasis on multi-modality. Note that the term ‘work environment’ is loosely defined in the paper but the approach is explained via a concrete example of a warehouse picking system. At design-time, developers can specify three versions (vocal, graphical, and multi-modal) of the concrete UI model using the UI definition language: MARIA \cite{b10}. Developers can also define a set of adaptation rules via an XML-based DSL to adapt the concrete UI models per given context-of-use (CoU) scenario changes. At run-time, the system listens to CoU changes and executes adaptation rules to tailor the UI. As for the limitations of the study, the developers are required to design three versions of the same UI model resulting in effort redundancy. The MDE process can be further refined to use a single abstract model to generate concrete UI model instances based on the modality mode. Another notable point is that the CoU management module is explained as a black box component in the study. Therefore, not enough information is presented to elaborate on the context-modelling capabilities of the adaptive UI system. Similarly, there is a lack of information on the metamodel of the adaptation rules DSL leading to vagueness regarding its UI adaptation capabilities.

    Minon et al., \cite{b11} propose a tool named Adaptation Integration System (AIS) that provides an adaptation rules integration extension for MDE tools that conform to the UI abstraction levels presented in the CAMELEON Reference Framework \cite{b3}. the purpose of the AIS is to automatically tailor the UIs to meet the UI accessibility requirements of user groups with special needs such as the visually, hearing, and cognitively impaired both at run-time and design-time. The AIS is also comprised of a compilation of UI adaptation rules catered for the above-mentioned user groups. At design-time, the designer of the MDE UI tool manually enters a UI model belonging to any of the CAMELEON framework’s abstraction levels along with parameters that indicate the user’s disability. The designer will then receive an adapted UI model according to the user's disability. To explain the run-time adaptation integration process between an MDE UI tool and the AIS at the Concrete UI (CUI) level, Minon et al. give an example: when the MDE UI tool detects a context change, the contextual events and the CUI model of the tool are sent to the AIS. Thereafter, AIS matches and selects suitable adaptation rules based on the request parameters and applies them to the CUI model. Finally, the Final UI generator of the MDE UI tool receives the adapted CUI model instead of the original model and generates an adapted Final UI \cite{b11}. A limitation of AIS is that the adaptation rules do not demonstrate the full coverage of accessibility requirements (i.e., navigational and timed activity adaptations) of WCAG.

\section{Research Gaps}\label{RG}

Each study reviewed in section \ref{RW} can address some accessibility issues of seniors through UI adaptations. However, our in-depth assessment of each approach revealed that there are also accessibility limitations in each of them, especially when compared side-by-side to a standardised set of guidelines such as WCAG. For example, \cite{b16}'s approach is one of the most recent and mature ones from section \ref{RW}, but it does not consider accessibility features such as time-sensitive activity adaptations (eg: modifying the duration of timed activities, timeouts, and re-authentications) or the support for user's language proficiency. Similarly, \cite{b11} also possesses issues such as the lack of support for navigational adaptations (eg: modifications to the navigational routes to bypass blocks and have multiple pathways). Table \ref{tab:gaps} summarises the major accessibility gaps that we have identified (in reference to WCAG) in the existing studies based on the information provided. Some studies do not contain enough information such as metamodels, in-depth explanations of methodology, or examples/prototypes to fully identify all accessibility gaps in them.

\vspace{-17pt}
\begin{table}[h]
\caption{\label{tab:gaps} Major accessibility feature gaps in Section \ref{RW}}
\vspace{-10pt}
\begin{center}
\begin{tabular}{|m{20em}|l|} 
\hline
    \textbf{Adaptive UI accessibility feature gaps} & \textbf{Paper} \\ 
    \hline 
         Mult-modality support (eg: text to speech, speech-to-text, and screen magnifiers) & \cite{b8, b16, b12, b13}\\[1ex] 
    \hline 
         UI navigation support (eg: bypassing block, alternative routes, and multiple routes) & \cite{b13, b11}\\[1ex] 
    \hline 
         Presentation/style/layout adaptation support (eg: border, margin, padding, and font) & \cite{b8}\\[1ex]
    \hline 
         Support for platform-related variability (eg: screen dimensions and device type) & \cite{b8}\\[1ex]
    \hline 
         Adaptations for time-sensitive UI activities (eg: timed activities, timeouts, and re-authentications) & All\\[1ex]
    \hline 
         Error handling and UI help & All except for \cite{b15}\\
    \hline 
         Support for language proficiency-related UI adaptations (eg: cross-language support, reading comprehension, idioms, and jargon) & All\\[1ex]
    \hline 
\end{tabular}
\end{center}
\end{table}
\vspace{-15pt}

Even approaches that are mature in modelling accessibility scenarios via UI adaptations, such as \cite{b16, b13, b7}, still contain serious limitations in terms of adaptive UI support for senior users. We believe that many consider the accessibility needs of their target end users in an ad-hoc manner without referencing a framework and thus leading to the lack of coverage in some areas. Furthermore, even the studies that mention the incorporation of accessibility standards, such as \cite{b7} and \cite{b11}, only do so in a limited number of cases such as the guidelines related to text size and background/foreground contrast. 

Thus, the developers of UI adaptation-capable applications need to consider an ‘accessibility first’ paradigm where the adaptive accessibility UI needs of end users are treated as first-class citizens in the application design process. Only then they would be able to provide full accessibility support for a user group similar to seniors whose age-related UI needs are diverse and variable.  

In order to achieve this goal, we can leverage some of the most promising approaches to MDE UI adaptation from the existing works. Based on our literature review, accessibility has been integrated into the presented approaches with the following two techniques. Firstly, accessibility situations a user may face during application usage are represented with Context-of-Use scenario models utilising the user, platform and environmental context dimensions. Secondly, UI adaptation rules that reference each CoU scenario are used to perform concrete UI adaptations. 

\section{A Research Roadmap}\label{RR}

In order to devise an accessibility-first approach to developing model-driven adaptive UIs, we believe that the following goals and objectives are to be followed as our ongoing and future work.

\begin{figure}[!ht]
\centerline{\includegraphics[scale=0.06]{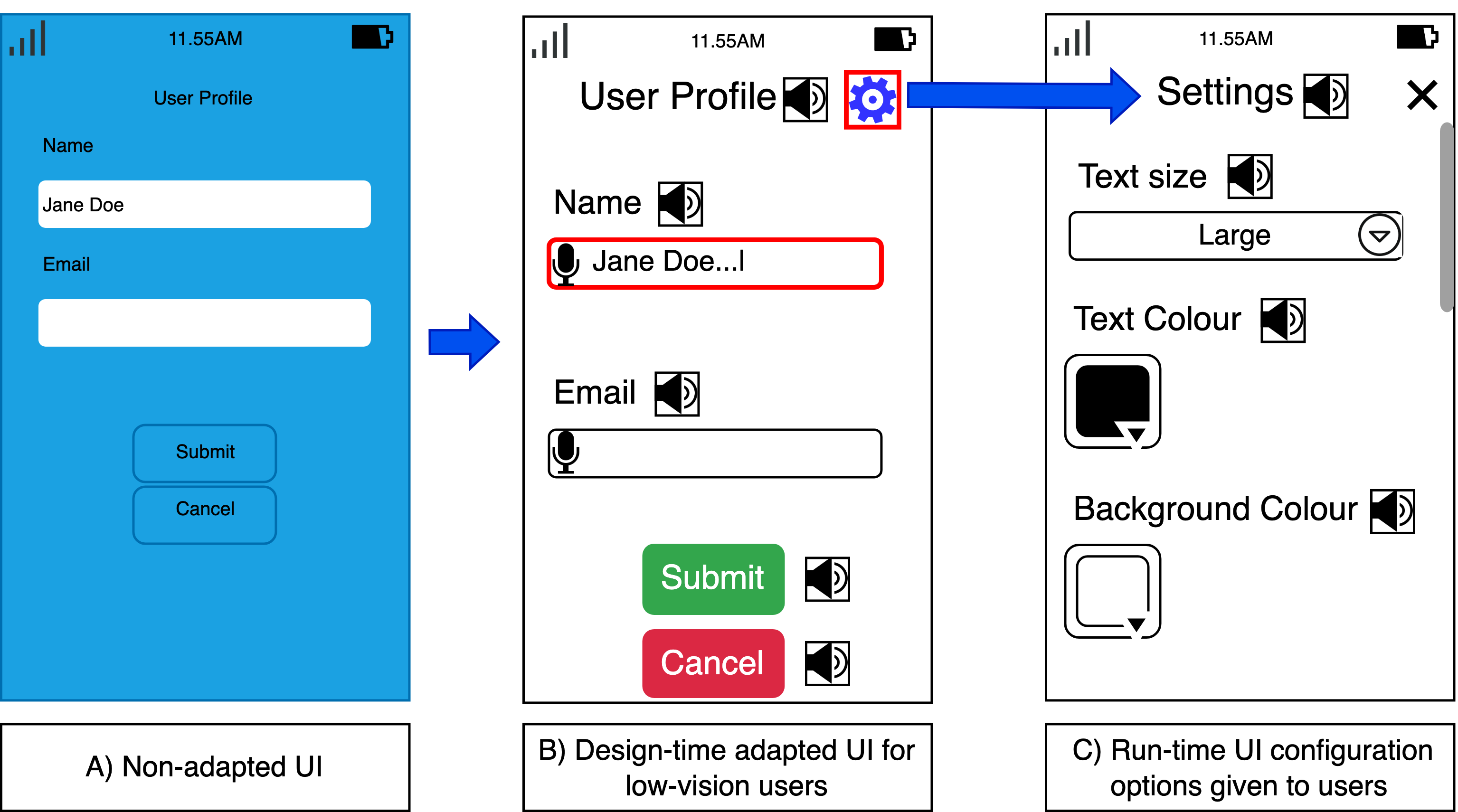}}
\vspace{-10pt}
\caption{A UI adaptation scenario for seniors with low-vision where a non-adapted UI (A) is being design-time (B)/run-time (C) adapted with UI modifications to contrast ratio, text-size, multi-modality (text-to-speech and speech-to-text)}
\label{fig2}
\end{figure}

\subsection{Our Goal}

An MDE-based adaptive UI modelling and code generation tool that aims to enable software developers to build accessible software UI that meets the age-related needs of seniors.

\subsection{Key Objectives}

\vspace{-15pt}

\begin{figure}[hbp]
\centerline{\includegraphics[scale=0.07]{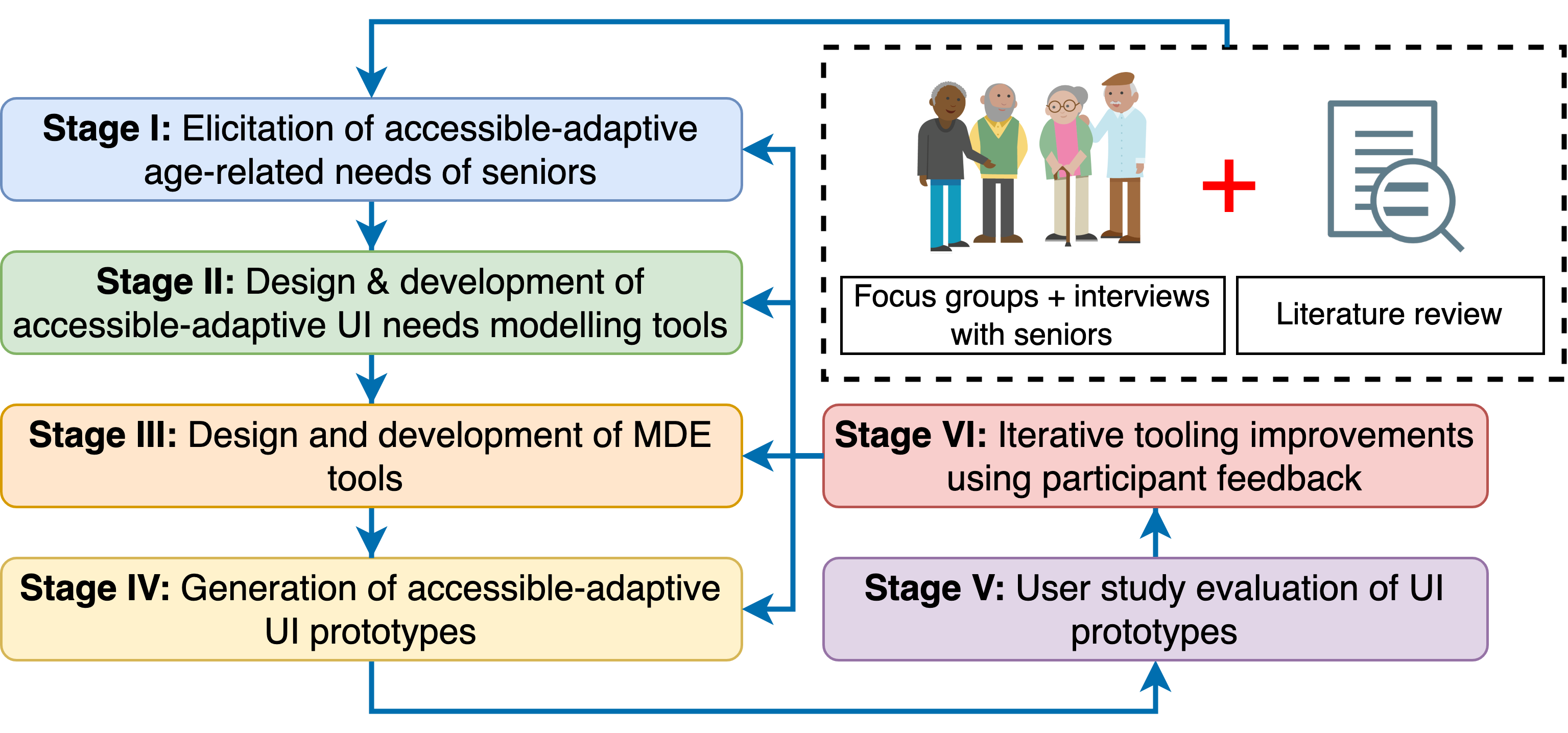}}
\vspace{-10pt}
\caption{The research roadmap}
\label{fig1}
\end{figure}


    \begin{enumerate}
        \item[\textbf{1)}] \textbf{How do we elicit a concrete set of accessibility and adaptive UI requirements from seniors?}
    \end{enumerate}

    Requirement elicitation occurs through two information sources: 1) \textbf{Focus group studies} and follow-up \textbf{interviews} with multiple groups of senior users; 2) Continued literature review on age-related UI accessibility needs. We are conducting both activities to generate a rich and complete set of requirements for adaptive, MDE-based UIs for senior end users.

    \begin{enumerate}
        \item[\textbf{2)}] \textbf{How do we model the accessibility needs of seniors in a WCAG-compliant manner?}
    \end{enumerate}

    We believe that two new WCAG-compliant \textbf{domain-specific languages (DSL)} are needed to model the requirements elicited from the previous step: 1) A graphical DSL for modelling context-of-use scenarios; 2) A textual DSL for specifying UI adaptation rules are needed. We take inspiration from \cite{b16}'s ContextML and AdaptML DSLs that serve a similar purpose. ‘WCAG compliance’ means that each user-platform-environment context-of-use combination that represents an accessibility scenario of the elicited requirement and the adaptation rules associated with them are capable of fulfilling WCAG success criteria. In case WCAG is not sufficient to cover all of the identified accessibility requirements, we also look at the existing accessibility literature, especially in the domain of Human-Computer Interactions (HCI).

    \begin{enumerate}
        \item[\textbf{3)}] \textbf{How do we design and develop an MDE process to generate Final UI?}
    \end{enumerate}

    We aim to integrate both the context-of-use model and adaptation rules model from step 2 into an MDE process so that the models can adapt a non-adapted UI model into an adapted UI model. Such an MDE process is to be capable of both: 1) \textbf{Design-time UI adaptation} - Developers can use the context-of-use information of end users and tailor the UI at design-time to suit the persona that the user belongs to; 2) \textbf{End-user can personalise the UI at run-time} - we acknowledge that developers are very different from seniors and hence, it is challenging for them to capture all the diverse age-related UI needs of seniors at design-time. Therefore, we plan to provide elderly users with the means to customise the UI according to their personal preferences.

    \begin{enumerate}
        \item[\textbf{4)}] \textbf{How do we generate prototype adaptive software UI for seniors?}
    \end{enumerate}

    The elicited requirements from step 1 are modelled via DSLs from step 2, and these models are then input into the MDE process from step 3 to generate prototype UIs. These are then capable of UI adaptation on both run-time and design-time to suit different senior end-user needs. An example accessibility scenario is illustrated in Figure \ref{fig2}.

    \begin{enumerate}
        \item[\textbf{5)}] \textbf{How do we evaluate prototypes with seniors?}
    \end{enumerate}

    A \textbf{user study} is required to be conducted to evaluate the user experience of our generated prototypes from the previous step, with  \textbf{senior users as study participants}.

    \begin{enumerate}
        \item[\textbf{6)}] \textbf{How do we use feedback from the evaluation?}
    \end{enumerate}

    Based on the feedback from these user studies, refinements will be made to the artefacts from objectives 1, 2, 3, and 4. Overall, \textbf{an iterative approach} is used to fine-tune our accessible adaptive UI generation tools, as outlined in Fig. \ref{fig1}.




\section{Conclusion}\label{CC}

The world's population is ageing rapidly, but software applications more often than not overlook the accessibility needs of seniors - thus hampering the user experience of senior users and subsequently inducing digital exclusion. As a solution, MDE-based UI adaptivity has been proposed and the literature contains mature methods that allow developers to model accessibility context-of-use scenarios and specify UI adaptation rules at design time. On run-time, depending on context change it is possible to adapt the UI in an automated manner or a user-configurable manner. However, these existing work lacks comprehensiveness when attempting to integrate accessibility into their methods and tooling and we believe that treating accessibility as a first-class citizen at the UI modelling stage is necessary. Our roadmap aims to use WCAG as a framework to design DSLs that are capable of modelling both CoU scenarios and UI adaptation rules so that for any given accessibility scenario, we can adapt the UI at both run-time and design-time while also meeting WCAG specifications. 

\section*{Acknowledgments}

Authors are supported by ARC Laureate Fellowship FL190100035.

\bibliographystyle{IEEEtran}
\bibliography{main}

\vspace{12pt}
\end{document}